\documentclass[aps,twocolumn,superscriptaddress,floatfix,prb]{revtex4-2}

\usepackage{lipsum}
\usepackage{graphicx}
\usepackage{amsmath,amssymb}
\usepackage{braket}
\usepackage{mathtools}
\usepackage[hidelinks]{hyperref}
\usepackage{multirow}
\usepackage{color}
\usepackage{bbm} 
\usepackage[normalem]{ulem}
\usepackage{amsfonts}
\usepackage{float}
\usepackage{pdfpages} 
\makeatletter
\usepackage{subfigure}
\usepackage{physics}

\def\bra#1{\mathinner{\langle{#1}|}}
\def\ket#1{\mathinner{|{#1}\rangle}}
\def\braket#1{\mathinner{\langle{#1}\rangle}}

\newcommand{\Ep}{\mathcal{E}}

\def\beq{\begin{equation}}
\def\eeq{\end{equation}}
\def\bea{\begin{eqnarray}}
\def\eea{\end{eqnarray}}

\usepackage{orcidlink}

\AtBeginDocument{\let\LS@rot\@undefined}
\makeatother

\begin{document}

\title{Distinguishability Transitions from Global Quantum Snapshots}

\newcommand{\unige}{
  Department of Theoretical Physics, University of Geneva,
  24 quai Ernest-Ansermet, 1211 Gen\`eve, Switzerland
}

\newcommand{\princeton}{Department of Electrical and Computer Engineering, Princeton University, Princeton, NJ 08544, USA}

\newcommand{\gqc}{
  Geneva Quantum Center, University of Geneva
}

\author{Catherine McCarthy \:\orcidlink{0009-0008-6588-670X}}
\affiliation{\unige}
\affiliation{\gqc}

\author{Sarang Gopalakrishnan \:\orcidlink{0000-0002-7493-7600}}
\affiliation{\princeton}

\author{Romain Vasseur \:\orcidlink{0000-0002-4636-4139}}
\affiliation{\unige}
\affiliation{\gqc}

\begin{abstract}

Quantum states generated by generic time evolution are locally featureless: local measurements return random outcomes that are identical for all states. However, \emph{global} snapshots in the computational basis are surprisingly effective at distinguishing between quantum states. We explore the ability of Bayesian classifiers to discriminate between many random quantum states using a small number of measurement outcomes. We identify a phase transition in the ability of the classifier to distinguish between $k$ candidate states with $m$ shots controlled by the variable $x = m / \ln k$, and derive the threshold value $x_c$.  We extend our results to low-depth circuits near the onset of anti-concentration, as well as to noisy circuits and phase-random states. 
%

\end{abstract}
\vspace{1cm}

\maketitle

\section{Introduction}

Generic quantum evolution hides information about initial states from local measurements through the process of thermalization \cite{deutsch1991quantum, srednicki1994chaos, rigol2008thermalization, dalessio2016quantum}. In systems with no conservation laws, such as random circuits, any initial state becomes locally indistinguishable from the maximally mixed state \cite{nahum2017quantum, nahum2018operator, vonkeyserlingk2018operator, fisher2023random}; in particular, \emph{all} initial states eventually look locally \emph{identical}. This indistinguishability goes beyond local measurements and arises at surprisingly early times, with recent work showing that even relatively shallow circuits (of depth slightly beyond logarithmic in the number of qubits, $N$) can generate states that are computationally indistinguishable from random states---even when the states are orthogonal in the many-body Hilbert space and there exists a measurement that can perfectly distinguish them in principle \cite{ji2018pseudorandom,brakerski2019pseudo,aaronson2022quantum,schuster2025random}. Thus, if one is restricted to $\mathrm{poly}(N)$ samples of a state and classical runtime, distinguishing between two randomly chosen quantum states, or between two states generated by deep quantum circuits, is presumably intractable. 

In this work we consider relaxing these restrictions so that quantum samples remain expensive but classical computation is cheap, while simultaneously requiring that all measurements must be performed in the computational basis.  Specifically, we assume that we have classical models of the quantum states of interest, $\ket{\psi_\alpha}$, which can be queried to give the probability $p(\vec{m}|\psi_\alpha) \equiv |\langle \vec m | \psi_\alpha \rangle|^2$ that a computational-basis measurement of the state $\ket{\psi_\alpha}$ will yield the bitstring $\vec m$. For a random pure state, it is well-known that these probabilities are not uniform across bit-strings --- instead, each state has its distinctive distribution of probabilities over bit-strings, featuring some outcomes that are likelier to occur (or ``heavier'') than others \cite{aaronson2016complexity}.  Since each bit-string is exponentially unlikely to be measured, estimating measurement probabilities is a task that is generally classically intractable but unavoidable in certain contexts, notably including benchmarking Random Circuit Sampling (RCS) experiments \cite{arute2019quantum}.   In these situations, the linear cross-entropy between the observed bit-string distribution and the theoretical model~\cite{boixo2018characterizing, arute2019quantum} is commonly used to detect the sampling of heavy bitstrings, which is classically hard \cite{aaronson2016complexity}; however, recent works have demonstrated that this quantity is vulnerable to classical spoofing attacks \cite{aaronson2020spoofing,barak2021spoofing,gao2024limitations}.

As an alternative probe of heavy bitstrings, one can consider tasks in which snapshots are ``classified'' according to the state they came from.  As a simple illustration of this strategy, consider an experiment that reliably generates one of two random states $\ket{\psi_\alpha}$, with $\alpha=0,1$, with equal prior probability. Measuring the unknown state in the computational basis yields a bit-string $\vec m$, and querying the classical models of the two states yields conditional probabilities $p(\vec m | \psi_\alpha)$ for $\alpha = 0, 1$. With access to classical models of both states, the optimal strategy for guessing the label $\alpha$ consists of applying Bayes' rule, which in this simple example is equivalent to picking the state $\alpha$ with the highest probability of producing the bit-string $\vec m$. A simple calculation \cite{vasseur2026leshoucheslecturesrandom} shows that if one were to repeat this game many times, with randomly chosen $\ket{\psi_0}, \ket{\psi_1}$, the strategy outlined above succeeds $3/4$ of the time.

This paper explores two generalizations of the state-distinguishing game introduced above. First, we consider the problem of distinguishing $k$ random states, given $m$ shots of the unknown state, chosen from \(k\) candidate states, in the limit where $k, m, N$ are all taken to be asymptotically large.  We find a phase transition in this problem when $m \sim \ln k$, indicating that $m$ measurements suffice to distinguish a given random state from ${\cal O}(\exp m)$ many others, consistent with recent work on state certification (see Appendix J of Ref.~\cite{huang2025certifying}).  After deriving the transition's threshold value, we discuss connections between the quantum distinguishability game setup and noisy classical channel coding.  We then extend our analysis beyond random states and consider phase-random states, states corrupted by noise, and states generated by short-depth quantum circuits.  We find that noise and finite depth have opposite effects on the success probability: noise suppresses it, by making the bit-string distribution essentially uniform, while short-depth circuits do not feature anti-concentration of the output distribution, so the heavy outputs are parametrically heavier than typical ones and the success probability converges to unity as $N \to \infty$.

The rest of this manuscript is organized as follows. Sec. \ref{sec:background} briefly introduces Bayesian classification and details the one-shot distinguishability game with two candidate states.  Then Sec. \ref{sec:k-m} discusses an extension of the distinguishability game to arbitrary numbers of candidate states and shots.  Finally, Sec. \ref{sec:extensions} generalizes the distinguishability games when the pure Haar-random assumption is relaxed for the candidate states --- here we discuss phase-random states, noisy states, and states prepared from finite-depth circuits.


\section{Bayesian classification and state distinguishability} \label{sec:background}

\subsection{Bayesian classification}

A foundational result in probability theory is \textit{Bayes' theorem}, which describes updating prior to posterior probabilities upon obtaining new information \cite{duda2001pattern}.  The prior probability $p(\alpha)$ denotes the probability of event $\alpha$ occurring without taking into account any other information.  If a new event $\beta$ occurs, the probability of event $\alpha$ must be updated to account for any statistical dependence between the two events.  The update rule is given by Bayes' theorem:
\begin{equation}
    p(\alpha| \beta) = \frac{p(\beta| \alpha) p(\alpha)}{p(\beta)},
\end{equation}
where $p(\alpha| \beta)$ is the conditional probability of $\alpha$ given the occurrence of the event $\beta$.  In the context of the update rule, $p(\alpha|\beta)$ is also known as the posterior probability.

A \textit{Bayesian classifier} is a decision rule used to predict which among a set of events $\alpha_j$ for $j=0,\dots,K$ will occur.  The classifier has access to the prior probabilities $p(\alpha_j)$; if the event $\beta$ occurs after the declaration of the prior probabilities, they may be updated according to Bayes' rule $p(\alpha_j | \beta) = \frac{p(\beta | \alpha_j) p(\alpha_j)}{\sum_i p(\beta| \alpha_i) p(\alpha_i)}$.  If asked to predict the true label $j_{\rm true}$, the optimal decision is to select the event $\alpha_j$ that has the highest posterior probability-- that is, the predicted label is given by $j_{\rm pred}=\mathrm{argmax}[p(\alpha_0 |\beta),\dots, p(\alpha_K| \beta)]$. 

One quantity of interest related to the classifier is the probability that it successfully predicts the true label $j_{\rm pred}=j_{\rm true}$.  Consider the scenario in which a label $j$ is chosen according to the distribution of prior probabilities $p(\alpha_j)$; after this choice, one of a set of events $\beta_k$ for $k=0,\dots,M$ occurs with known conditional probabilities $p(\beta_k| \alpha_j)$.  The classifier's success probability averaged over all possible events $\beta_k$ is given by the following expression:
\begin{equation}
    \begin{aligned}
        P_s = \sum_{j, k} \Big[&p(\alpha_j) p(\beta_k| \alpha_j) \cdot \\& \prod_{i \neq j}\Theta\big(p(\alpha_j)p(\beta_k| \alpha_j) - p(\alpha_i)p(\beta_k| \alpha_i) \big)\Big],
    \end{aligned}
\end{equation}
where $\Theta(\cdot)$ is the Heaviside function, with  $\Theta(0)=1/2$ to account for binary ties. (In all problems with
more than two candidate labels considered in this manuscript, the largest
posterior probability is unique almost surely.)  In this manuscript, we will be interested in the average success probability $P_s$ of such an optimal Bayesian classifier predicting the correct label in the context of quantum state discrimination conditioned on the outcome of a measurement.  In keeping with the jargon of the field, we will refer to the Bayesian classifier as our \textit{decoder}.

\subsection{The single-shot distinguishability game} \label{subsec:oneshot}

\begin{figure*}[ht!]
    \centering
    \includegraphics[width=0.5\linewidth]{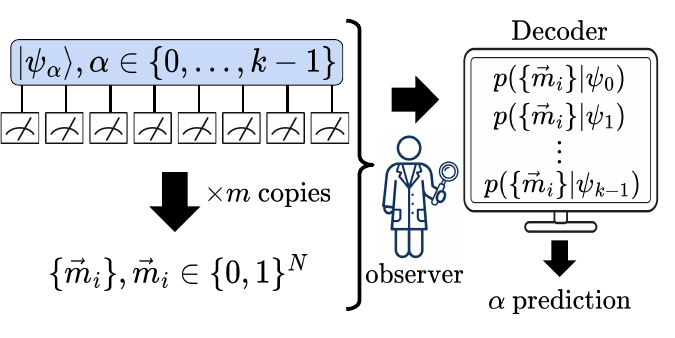}
    \includegraphics[width=0.4\linewidth]{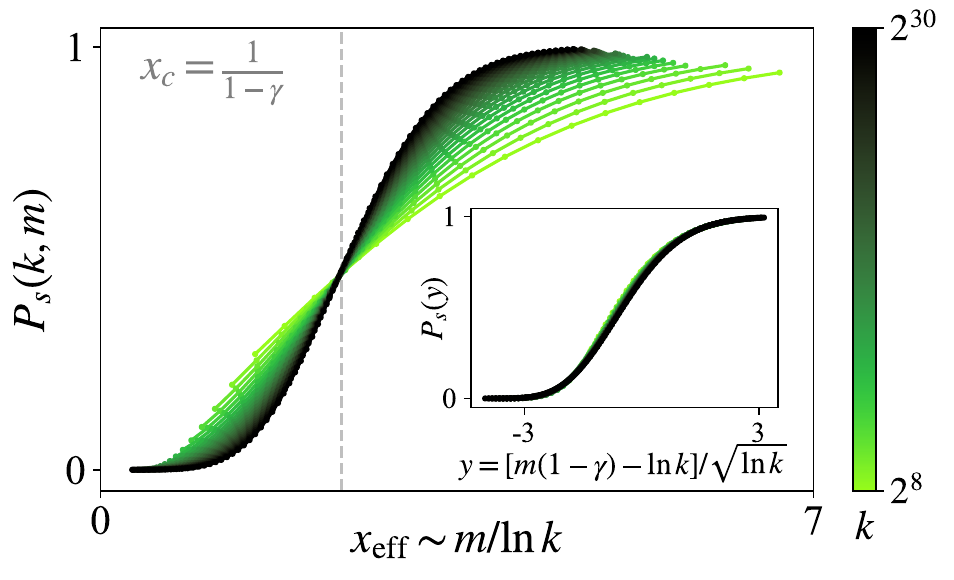}
    \hfill
    \caption{Left: Protocol for distinguishing $k$ random states with $m$ global snapshots.  In this setup, an observer is presented with $m$ copies of a random state $\ket{\psi_\alpha}$ with $\alpha \in \{0, \dots, k-1\}$ and provided with exact knowledge of all $k$ states.  After measuring each site of each of the $m$ copies in the computational basis to obtain a set of bitstrings $\{\vec m_i\}$, the observer queries a decoder in order to guess the label $\alpha$. Here we take the decoder to be an optimal Bayesian classifier that outputs the label $\alpha$ associated with the largest probability $p(\{\vec m_i\} | \psi_\alpha)$, where $p(\{\vec m_i\} | \psi_j)$ is computed for all $j \in \{0, \dots,k-1\}$ from the Born probabilities associated with the bitstrings $\{ \vec m_i\}$.  Right:  Decoder success probability $P_s(x_{\rm eff})$, where $x_{\rm eff} = m/(\ln k - \frac{1}{2} \ln [2 \pi m(\pi^2/6-1)])$ for various values of $k$ and $m$ obtained from numerical integration of Eq.~\eqref{eq:km}.  At the threshold value $x_c = 1/(1-\gamma)$ the decoder undergoes a transition from a phase in which the decoder cannot distinguish between the $k$ states to one in which the decoder can successfully predict $\alpha$.  Inset: accounting for fluctuations of $\sqrt m \sim \sqrt{\ln k}$, the data  collapses in terms of the variable $y = [m(1-\gamma) - \ln k]/\sqrt{\ln k}$.
    }
    \label{fig:fig1}
\end{figure*}

Our primary interest is studying state distinguishability as a Bayesian classification task.  For the basic setup, consider a game in which an observer is presented with an $N$-qubit random state $\ket{\psi_\alpha} = \sum_{\vec s \in \{0,1\}^N} c_{\vec s, \alpha} \ket{\vec s}$ drawn from the Haar measure. The observer is promised that the label of this state is either $\alpha=0,1$ and provided with both sets of coefficients $\{c_{\vec s, \alpha}\}$.  The observer is then tasked with guessing the label $\alpha$ of the state. In this game, both states have equal prior probabilities $p(\alpha=0) = p(\alpha=1) = \frac{1}{2}$; therefore, the observer has an initial success probability $P_s=1/2$ from guessing the label $\alpha$ at random.  

The observer may increase their success probability by performing measurements in order to gain additional information about the state.  The obvious strategy is to perform an optimal Helstrom measurement, which is the projective measurement that minimizes the average probability of error in discriminating between two (not necessarily orthogonal) quantum states \cite{helstrom1969quantum}.  In this work, we instead consider the scenario where the observer is restricted to computational basis measurements.  The best strategy is to perform the measurement and choose the label $\alpha$ that corresponds to the larger Born probability of the measured bitstring $\vec m$.  After measurement, the probabilities for each label $\alpha$ are updated according to Bayes' theorem $p( \psi_\alpha | \vec m) = \frac{p(\vec m | \psi_\alpha)}{p(\vec m|\psi_\alpha) + p(\vec m | \psi_{\alpha \oplus 1})}$, where here and throughout this manuscript, we use $\oplus$ to denote addition modulo 2.   The observer's success probability may then be written as the following:
\begin{equation}
    \begin{aligned}
        P_s &= \sum_{\alpha=0,1} \sum_{\vec m \in \{0,1\}^N} p(\psi_\alpha) p(\vec m| \psi_\alpha ) \Theta(p(\vec m| \psi_\alpha) - p(\vec m | \psi_{\alpha \oplus 1})) 
     \\&= \frac{1}{2} \sum_{\vec m \in \{0,1\}^N} \max[p(\vec m | \psi_0), p(\vec m| \psi_1) ].
    \end{aligned}
    \label{eq:max}
\end{equation}
\textit{A priori}, it is not intuitive that the observer can significantly increase their success probability above $P_s=1/2$ for a random state with access to only one shot; however, the simple example of a single random qubit illustrates the contrary.  Given $\ket{\psi_\alpha} = c_{\vec 0, \alpha}\ket{0} + c_{\vec 1, \alpha} \ket{1}$, with Born probabilities $p(\vec m| \psi_\alpha) = |c_{\vec m, \alpha}|^2$ sampled from the uniform distribution on $[0,1]$, the sum in eq.~\eqref{eq:max} may be easily evaluated as an integral $P_s(N=1) = \int_0^1 dw \int_0^1 dv ~\max(w,v) = \frac{2}{3}$,
for $w,v \sim p(\vec m| \psi_\alpha)$.  Note that in this expression, we are implicitly assuming that the success probability is averaged over different initial state $\alpha=0,1$ in order to faithfully sample the Born probability distribution. Extending this game to Haar-random $N$-qubit states requires drawing the Born probabilities from a Beta distribution $P(w)= (D-1)(1-w)^{D-2}$ \cite{bengtsson2007geometry}, where $D=2^N$. Eq.~\eqref{eq:max} may then be evaluated exactly for arbitrary $N$:
\begin{equation}
    \begin{aligned}
        P_s(N) &= D(D-1)^2\int_0^1dw ~w(1-w)^{D-2} \int_0^w dv ~(1-v)^{D-2},
    \end{aligned}
    \label{eq:arbitrary-N}
\end{equation}
which evaluates to $P_s(N)=\frac{3D-2}{4D-2}$.  Expanding this result in terms of a small $1/D$ parameter, we obtain:
\begin{equation}
    P_s(N \gg 1 ) = \frac{3}{4}\left(1-\frac{1}{6D} + \mathcal{O}(1/D^2)\right).
    \label{eq:large-limit}
\end{equation}
Taking the $N \rightarrow \infty$ limit gives $P_s \rightarrow 3/4$, which may also be recovered by directly evaluating Eq.~\eqref{eq:max} for the \textit{Porter-Thomas} distribution $P(w) = De^{-Dw}$ \cite{vasseur2026leshoucheslecturesrandom}.  As evident from the $\mathcal{O}(1/2^N)$ corrections to the Porter-Thomas result shown in Eq.~\eqref{eq:large-limit}, the decoder rapidly achieves a success probability $P_s \approx 3/4$ for just a few qubits.  Note that in the $N \rightarrow \infty$ limit, sampling the Born probabilities encoded in any particular Haar-random state amounts to faithfully sampling the Porter-Thomas distribution-- a consequence of this is that it is not necessary to implicitly average over different instances of $\ket{\psi_0}, \ket{\psi_1}$ in Eq. \ref{eq:large-limit}.

Given that the observer was only provided with a single copy of the state and was restricted to a fixed measurement basis, it may be surprising that the Bayesian classifier has a relatively high success probability $P_s=3/4$ in the large-$D$ limit.  For the Porter-Thomas distribution, large Born probabilities are suppressed but bitstrings associated with these rare relatively-large Born probabilities have an outsized likelihood of being measured; in other words, the \textit{measured} probabilities of the wrong label are drawn from the Porter-Thomas distribution $P(u) = e^{-u}$, where $u=Dw$, while the measured probabilities of the correct label are instead sampled from $P_{\rm corr}(u) = u e^{-u}$.  This bias toward high-probability outcomes is the key mechanism underpinning the decoder's enhanced success probability. It also underlies heavy output generation, which plays a key role
in arguments for the classical hardness of quantum
sampling~\cite{aaronson2016complexity}. The two relevant distributions are related through reweighting by the Born probabilities themselves as $P_{\rm corr}(u) = u P(u)$.  This reweighting is what sets the amount of information contained in a single snapshot, which can be quantified with the Kullback-Leibler (KL) divergence (or relative entropy) $D_{\rm KL}(\,P_{\rm corr}(u)||P(u)\,)= \mathbb E_{P_{\rm corr}}[\ln(P_{\rm corr}(u)/P(u))]$; here, we have $D_{\rm KL}(\,P_{\rm corr}(u)||P(u)\,)= \mathbb E_{P_{\rm corr}}[\ln u] = 1-\gamma$, where $\gamma$ is Euler's constant.

\section{Distinguishability transition} \label{sec:k-m}

We now consider variants of the single-shot distinguishability game.  There are multiple parameters that may be tuned to adjust the difficulty of the game.  For example, taking $m>1$ copies of the state allows the observer to take more global snapshots of the state and obtain the set of multiple output bitstrings  $\{\vec m_i\}$, with $i=0,\dots,m-1$, thus improving their guess of the label $\alpha$.  As the observer is provided with more shots, the success probability of their guess of the correct label approaches one exponentially.  Other generalizations of the single-shot distinguishability game will instead lower the decoder's success probability --- for instance, a game in which the observer must distinguish between $k>2$ Haar-random candidate states labeled by $\alpha = 0, \dots, k-1$ with access to only a single shot ($m=1$) becomes significantly more challenging.  Each label is associated with a prior probability $1/k$: while performing a measurement will increase the observer's success probability, in the limit of large $k$ the game becomes nearly impossible with $P_s(k \gg 1) \rightarrow 0$.  


From these considerations, the question of the difficulty of the game as $m$ and $k$ are tuned together arises naturally.  After measuring each of the $m$ copies to obtain the set of bitstrings $\{ \vec m_i \}$, the posterior probabilities $p(\psi_\alpha|\{\vec m_i\}) = \frac{\prod_i p(\vec m_i | \psi_\alpha)}{\sum_{\alpha'} \prod_i p(\vec m_i| \psi_{\alpha'})}$ may be used to predict the label $\alpha_{\rm pred} = \mathrm{argmax}[p(\psi_0| \{\vec m_i\}), \dots, p(\psi_{k-1}|\{\vec m_i\})]$.  Note that the $N \rightarrow \infty$ limit is taken first so that the rescaled Born probabilities $u=Dp(\vec m| \psi_\alpha)$ of each state are faithfully sampled from the Porter-Thomas distribution $P(u) = e^{-u}$, after which the limits of large $k$ and $m$ may be taken.  The average success probability of such a decoder may be written as 
\begin{equation}
    P_s(k,m)=\mathbb{E}_{S_m}[\big(1-Q(S_m)\big)^{k-1}],
    \label{eq:km}
\end{equation}
where $S_m = \sum_{i=0}^{m-1} \ln(\,Dp(\vec m_i|\psi_\alpha)\,)$ is the ``score'' that the classifier associates with the correct label and $Q(S_m)$ is the probability that the score of an incorrect candidate state is larger than $S_m$.  The independence in choosing the $k$ candidate states implies that the pairwise success probability $1-Q(S_m)$ is the same on average for each ``matchup'' between the correct state and the other $k-1$ candidate states.  While Eq.~\eqref{eq:km} does not have a closed form for arbitrary $k,m$, $P_s(k,m)$ may be evaluated in the limit where both parameters are asymptotically large.   The correct score $S_m$ approaches its typical value $\overline S_m$ in the large-$m$ limit; for rescaled Born probabilities of the correct label $u=Dp(\vec m_i | \psi_\alpha)$ sampled from the distribution $P_{\rm corr}(u) = u e^{-u}$, we obtain $\overline S_m = m(1-\gamma)$, where $\gamma$ is Euler's constant.  Evaluating the misclassification probability at the typical correct score value $Q(\overline S_m)$ yields
\begin{equation} \label{eqMis}
    Q(\overline S_m) \simeq \frac{e^{-m(1-\gamma)}}{\sqrt{2 \pi m(\pi^2/6-1)}}.
\end{equation}  
The exponential decay follows from a saddle-point
evaluation of the tail of the incorrect-label score.
Define $\Lambda(\lambda)=\ln\mathbb{E}_{P}[u^\lambda]
=\ln\Gamma(1+\lambda)$, the cumulant generating function of
$\ln u$ under $P(u)=e^{-u}$. The decay rate at the typical
correct score is
$\max_{\lambda}\big[\lambda\overline{S}_m/m-\Lambda(\lambda)\big]$.
The maximum is attained at $\lambda^\star=1$ and equals
$1-\gamma$.

This scaling leads to a threshold phase transition in this problem. To locate it, the average in Eq.~\eqref{eq:km} may be approximated by evaluating $P_s(k,m) \approx (1-Q(\overline S_m))^{k-1}$ at the typical score $\overline S_m$, which gives the following (approximate) expression in the limit of large $k$ and $m$:
\begin{equation}
    P_s(k,m) \simeq \exp \Big[-k \frac{e^{-m(1-\gamma)}}{\sqrt{2 \pi m(\pi^2/6-1)}}\Big].
    \label{eq:threshold}
\end{equation}
The success probability exhibits a threshold when $kQ(\overline S_m) \sim O(1)$, which occurs when $\ln k = m(1-\gamma) + \frac{1}{2}\ln[2 \pi m (\pi^2/6-1)]$.  To leading order, the threshold value is given by $x_c=1/(1-\gamma)$, where $x = m/\ln k$ is the appropriate scaling variable.  In practice, the subleading term decays extremely slowly for finite $k$ as $\ln(\ln k)/\ln k$, and so the threshold $x_c=1/(1-\gamma)$ is visible in an effective scaling variable that absorbs the subleading term, $x_{\rm eff} =m/\big(\ln k-\frac{1}{2}\ln[2 \pi m (\pi^2/6-1)] \big)$ (Fig. \ref{fig:fig1}).  By evaluating at $\overline S_m$, Eq.~\eqref{eq:threshold} locates the threshold value correctly but discards fluctuations about the typical score-- accounting for $\sqrt m \sim \sqrt{\ln k}$ fluctuations neatly collapses the numerically-obtained curves (Fig. \ref{fig:fig1} inset).  

Before proceeding, we offer a couple of comments on the threshold calculation.  The typical score computed above is exactly $\overline S_m = m D_{\rm KL}(\,P_{\rm corr}(u) || P(u) \,)$ and represents the excess information from sampling from the distribution of correct labels.  The threshold value $x_c=1/(1-\gamma)$ is then entirely determined by the KL divergence between the two distributions. Additionally, we note that the only limitation on the largeness of the parameters is that for a fixed state, the measurement outcomes are independent; however, after averaging over Haar-random states, repeated bitstrings reuse the same Born probability and introduce correlations. Such collisions are negligible when $m\ll\sqrt D$, by the birthday-paradox argument.  Therefore, another interesting consequence of the threshold calculation is that in the scaling limit of large $k$ with $m \sim\ln k$, it is possible to distinguish a single state from a set of $k \approx 2^N$ states that approximately spans the Hilbert space with only $m \sim N$ shots.

The leading-order $m \sim \ln k$ behavior strikingly indicates that with access to only a handful of shots, an observer may reliably distinguish between an exponential number of states.  In the context of distinguishing random quantum states, this scaling may initially seem counterintuitive; however, the same scaling is known to arise in the closely-related context of quantum hypothesis learning (see Appendix J of Ref. \cite{huang2025certifying}).  Furthermore, the exponential scaling in the number of shots also arises in the context of noisy classical channels \cite{cover2006elements}.  Consider a noisy classical channel that takes symbols $a \in \mathcal{A}$ as input and produces symbols $b \in \mathcal{B}$ as output, for two alphabets $\mathcal A, \mathcal B$.  A $(k,m)$ code is a defined procedure for transmitting one of $k$ messages from a set indexed by $\{0,\dots k-1\}$ through the channel that consists of both a procedure to encode each message $\alpha$ into a distinct codeword $a^{m}(\alpha)$ and a decoding procedure for the recipient to associate each $b^m$ with a particular $a^m$.  Since the channel can only send symbols $a \in \mathcal A$, transmitting a codeword $a^m$ effectively corresponds to using the channel $m$ times.  The rate of a $(k,m)$ code $R = \log_2(k)/m$ then gives the amount of information transmitted per use of the channel.  It is well-established~\cite{shannon1948, cover2006elements}  that there exists a value $C$, known as the
\textit{channel capacity}, below which reliable transmission is possible: for any
$R < C$, there exist $(k,m)$ codes with $k = 2^{mR}$ whose error probability approaches
zero as $m \to \infty$.
Closely related decodability transitions arise in spin-glass formulations of error-correcting codes. For capacity-achieving random code ensembles under optimal decoding, the information-theoretic threshold occurs at $R_c=C$~\cite{sourlas1989spin, vicente1999finite, mezard2009information}.  Therefore, the $m \sim \ln k$ scaling is well-established in the classical channel coding context.  

The $(k,m)$ state distinguishability game may be understood in direct analogy with the decodability transition in classical channels.  The sent message corresponds to the label $\alpha$ of one of the $k$ candidate states, and each of the $m$ shots counts as a single use of the channel in order to build up a transmitted codeword $b^m$.  Here, the (admittedly unusual) procedure to encode the message $\alpha$ is the preparation of the state $\ket{\psi_\alpha}$, while the optimal Bayesian classifier serves as the decoder.  Therefore, while perhaps surprising in the context of quantum state distinguishability, the mechanism underpinning the exponential scaling of transmitted information is well-grounded in standard information theory.

\section{State distinguishability without exact Haar-randomness} \label{sec:extensions}

The $k, m$ distinguishability game discussed above assumes that states provided to the observer are pure and Haar-random.  In practice, this is a fairly strong assumption due to the difficulty of preparing such states and the accumulation of errors.  Below we examine the breakdown of decodability as the noiseless Haar-random assumption is loosened.  For concreteness, we mostly limit the discussion to the single-shot ($k=2, m=1$) game outlined in Sec.~\ref{subsec:oneshot}, although
the threshold calculation of the previous section carries over
unchanged to any ensemble for which the decoder is supplied with the exact model of the candidate states: the threshold is then
$x_c = 1/D_{\rm KL}(\,P_{\rm corr}(u)||P(u)\,)$, where $P(u)$ is the distribution of rescaled Born probabilities of a wrong label and $P_{\rm corr}(u) = u P(u)$ that of the correct one. (The noisy classifiers of Sec.~\ref{subsection:noise}  below are an exception, since there the decoder scores with a model that differs from the distribution the outcomes are drawn from.)

\subsection{Phase-random states} \label{subsection:phase-random}

Since all prior results relied on the Porter-Thomas distribution, it is natural to ask whether exact Porter-Thomas statistics are necessary to achieve the enhanced decoder success probability or whether sampling from a distribution with heavy-output generation is sufficient.  To address this point, we consider a variant of the single-shot distinguishability game with two pure phase-random states \cite{nakata2012phase,nakata2014generating}.  For $\alpha=0,1$, the two candidate states are given by $\ket{\psi_\alpha} = \frac{1}{\sqrt D}\sum_{\vec s} e^{i \theta_{\vec s,\alpha}}\ket{\vec{s}}$ with $\alpha=0,1$ and random $\theta_{\vec s,\alpha} \in [0,2\pi)$.  This class of phase-random states trivially constitutes a state $1$--design with $p(\vec m | \psi_\alpha) = 1/D~\forall \vec m, \alpha$ \cite{nakata2014generating}.

While the observer gains nothing from directly performing $Z$-measurements on $\ket{\psi_\alpha}$, further information may be extracted if the state is measured in a new basis.  For a basis defined by the rotation $U = \bigotimes_j e^{-i \epsilon X_j}$ for a fixed constant $\epsilon$, we find numerically that the success probability approaches the Haar-random value $P_s=3/4$ for sufficiently large $\epsilon$ as a function of the rescaled variable $\epsilon N^{1/2}$ (see Appendix \ref{app:phase-random} for numerical data and analytical explanation for the scaling variable).  From this we conclude that exact Porter-Thomas statistics are not necessary in order to have a high success probability, provided that the Born probabilities look sufficiently randomly-distributed in the basis of measurement.

\subsection{Noisy classifiers} \label{subsection:noise}

\begin{figure}
    \centering
    \includegraphics[width=0.85\linewidth]{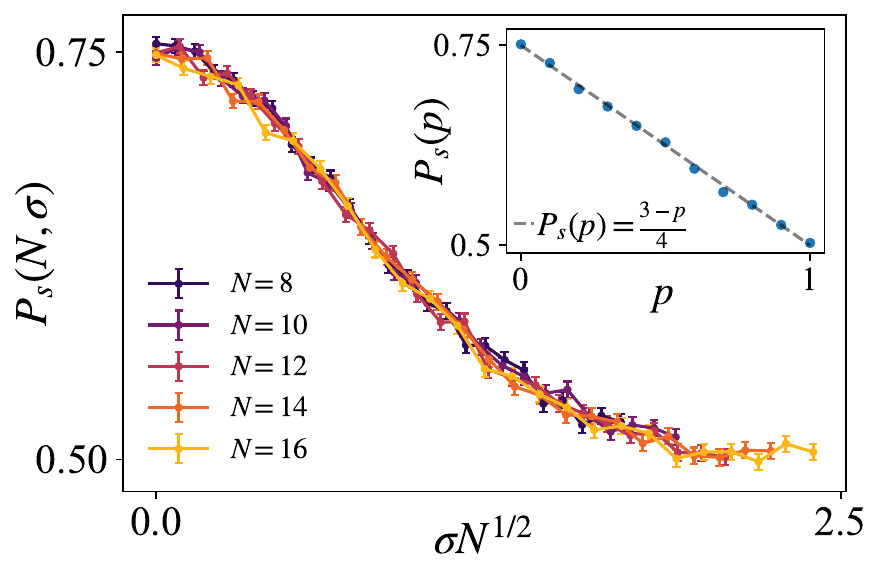}
    \caption{Success probability $P_s(\sigma, N)$ of decoding an $N$-qubit random state subject to random coherent errors acting on each site $j$ as $U_j = e^{-i \epsilon_j X_j}$, where $\epsilon_j$ are independently drawn from a Gaussian distribution of variance $\sigma^2$.  The success probability decays towards $1/2$ for large values of  $\sigma N^{1/2}$.  Inset: for a state under a global depolarizing channel of strength $p$, the decoder success probability takes the form $P_s(p) = (3-p)/4$.}
    \label{fig:noise}
\end{figure}

The original decoding protocol assumes that the observer has perfect knowledge of the states $\ket{\psi_\alpha}$ for $\alpha=0,1$.  We now extend our analysis to the situation in which the observer has imperfect knowledge of the state and study the breakdown of decodability.

\subsubsection{Random coherent error channel} \label{subsec:coh}

First we consider the case where the observer is given a noisy copy of the state $\ket{\psi'_\alpha}$ which has random coherent bit-flip errors $U_j = e^{-i \epsilon_j X_j}$ applied to each site $j$ (note phase errors do not affect the Born probabilities in the computational basis and thus have no impact on decodability).  The strength of the coherent error $\epsilon_j$ is chosen at random from a zero-mean normal distribution with variance $\sigma^2$.  Although the true Born probabilities in the computational basis are given by $p'(\vec m| \psi_\alpha) = |\braket{\vec m|U|\psi_\alpha}|^2$, the observer still uses the original probabilities $p(\vec m| \psi_\alpha) = |\braket{\vec m| \psi_\alpha}|^2$ to decode.  After averaging over the coherent errors, the Born probabilities
in the computational basis satisfy:
\begin{equation}
    \begin{aligned}
        p'(\vec m | \psi_\alpha) = \sum_{\vec s } c_s ~| \braket{\vec m \oplus \vec{s}|\psi_\alpha}|^2,
    \end{aligned}
    \label{eq:noisyaveraged-main}
\end{equation}
where $c_s = ( \frac{1 + e^{-2 \sigma^2}}{2} )^{N-|\vec s|}( \frac{1 - e^{-2 \sigma^2}}{2} )^{|\vec s|}$.  A short calculation (see Appendix \ref{app:coherent}) leads to the following expression for the decoder success probability in the large $D$ limit:
\begin{equation}
    P_s(\sigma)  = \frac{1}{2} + \frac{1}{4}\Big( \frac{1 + e^{-2 \sigma^2}}{2} \Big)^N.
\end{equation}
The success probability decays towards the pre-measurement value $P_s=1/2$ as the error strength grows, which is approached as $ P_s(\sigma,N) - \frac{1}{2} \sim e^{-\sigma^2 N}$ in the scaling regime $\sigma \ll1 $ with fixed $\sigma N^{1/2}$, consistent with the numerical collapse shown in Fig. \ref{fig:noise}.

\subsubsection{Local bit-flip noise} \label{subsec:bitflip}

We may also consider the scenario in which the observer is given a state $\rho_\alpha' = \Ep(\rho_\alpha)$, where $\rho_\alpha = \ketbra{\psi_\alpha}{\psi_{\alpha}}$ and $\Ep(\rho)$ is a noisy channel.  Since phase-flip noise will again not impact the Born probabilities, we study a local bit-flip channel $\Ep(\rho) = \sum_{\vec s \in \{0,1\}^N} p^{|\vec s| }(1-p)^{N-|\vec s| } X^{s} \rho X^s$, where $|\vec s|$ is the Hamming weight of bitstring $\vec s $ and $X^s = \bigotimes_j X_j^{s_j}$.  We find the same results as the random coherent error channel with $p \sim \sigma^2$.  

The scaling of the decoder accuracy may be analytically understood by studying the probabilities in the presence of noise of strength $p$.  It can be shown that $p(\vec m | \Ep(\rho_\alpha))$ may be written in the following form:
\begin{equation}
    \begin{aligned}
        p(&\vec m | \Ep(\rho_\alpha)) =  \sum_{\vec s \in \{0,1\}^N} p^{|\vec s|}(1-p)^{N-|\vec s|} ~|\braket{\vec m \oplus \vec s | \psi_\alpha}|^2. \\
    \end{aligned}
    \label{eq:bitflip}
\end{equation}
The form of eq.~\eqref{eq:bitflip} is identical to the probabilities averaged over coherent bit-flip errors (eq. \ref{eq:noisyaveraged-main}) when identifying $p = \frac{1}{2}(1 - e^{-2 \sigma^2})$.  For small values of $\sigma^2$, we then have:
\begin{equation}
    p \approx \frac{1}{2}\big( 1 - (1 - 2\sigma^2 + \mathcal{O}(\sigma^4) ) \big) \sim \sigma^2,
\end{equation}
which yields $P_s - 1/2 \sim e^{-Np}$.  Note that despite the similarities between the random coherent error channel and the noisy bit-flip channel, the mechanism that leads to reduced decodability is different in each case-- in contrast to the noisy bit-flip channel, which fully erases information, the observer could recover $P_s=3/4$ in the coherent error case if provided with knowledge of the errors accrued on each site.

\subsubsection{Global depolarizing channel}

We additionally study the global depolarizing channel $\Ep(\rho) = (1-p)\rho + p \mathbbm{1}/D$.  In the $p =0$ limit, we recover the Porter-Thomas result $P_s=3/4$, while in the $p=1$ limit the density matrix is replaced with the maximally mixed state $\mathbbm{1}/D$ and no additional information is obtained from measurements, yielding $P_s=1/2$.  

For intermediate noise of strength $p$, the noisy probabilities $p'(\vec m| \psi_\alpha) = (1-p) p(\vec m|\psi_\alpha) + p/D$ are given by directly interpolating between the $p=0$ and $p=1$ limits.  Therefore the success probability is given by:
\begin{equation}
    P_s(p) = \frac{1}{2} \sum_{\vec m, \alpha} \big[(1-p) p(\vec m|\psi_\alpha) + \frac{p}{D} \big] \Theta \big( p(\vec m| \psi_\alpha) - p (\vec m | \psi_{\alpha \oplus 1}) \big).
\end{equation}
The first term in the brackets yields the $3/4$ from the standard single-shot distinguishability game weighted by a factor of $1-p$, while the second term evaluates to $p/2$.  Therefore, the full expression for the noisy success probability evaluates to $P_s(p)=\frac{3-p}{4}$, as confirmed by numerical data (Fig. \ref{fig:noise}).

\subsubsection{Noisy thresholds}

We now extend the analysis of the threshold of the $(k,m)$ distinguishability game of Sec.~\ref{sec:k-m} to all three noise channels, retaining a decoder that scores measurement outcomes using the ideal candidate probabilities.  The distribution of rescaled Born probabilities $u=Dp(\vec m| \psi_\alpha)$ used by the decoder is a mixture of the ideal case, where correct labels are drawn from $P_{\rm corr}(u) = ue^{-u}$, and the case where the correct labels are independent Porter-Thomas-distributed variables with $P(u)=e^{-u}$.  Let $c$ be the probability that the outcome is drawn from the ideal distribution-- for the coherent, bit-flip, and global depolarizing channels, we have $c = \big( \frac{1 + e^{-2\sigma^2}}{2} \big)^N$, $(1-p)^N$ and $1-p$, respectively.  Note that as in Sec. \ref{sec:k-m}, we first take the large-$D$ limit at fixed $c\in(0,1]$, followed by the large-$k,m$ limit.  For a measured bitstring $\vec m$, the rescaled ideal Born probability $u=Dp(\vec m|\psi_\alpha)$ of the correct candidate $\alpha$ is therefore distributed as $q_c(u) = c\, u e^{-u} + (1-c)\, e^{-u}$, which reproduces all three single-shot results as $P_s = \mathbb E_{q_c}[\,1-e^{-u}\,] = 1/2 + c/4$.  For all three noise channels, the incorrect labels are still scored against a random Porter-Thomas variable drawn from $P(u) = e^{-u}$.  The typical score is lowered from the ideal value of $m(1-\gamma)$ to $\overline S_m =m \mathbb E_{q_c}[\ln u] = m(c - \gamma)$; the saddle point that led to Eq.~\eqref{eqMis} then shifts to $\Lambda'(\lambda^\star) = c-\gamma$, with $\Lambda(\lambda) = \ln \Gamma(1+\lambda)$. The modified threshold is therefore given by:
\begin{equation}
    x_c(c) = \frac{1}{ \lambda^\star (c-\gamma) - \Lambda(\lambda^\star)},
    \label{eq:noisythreshold}
\end{equation}
recovering $1/(1-\gamma)$ at $c=1$.  

\subsection{Probing anti-concentration in finite-depth circuits} \label{subsec:anti-concentration}

\begin{figure}
    \centering
    \includegraphics[width=0.85\linewidth]{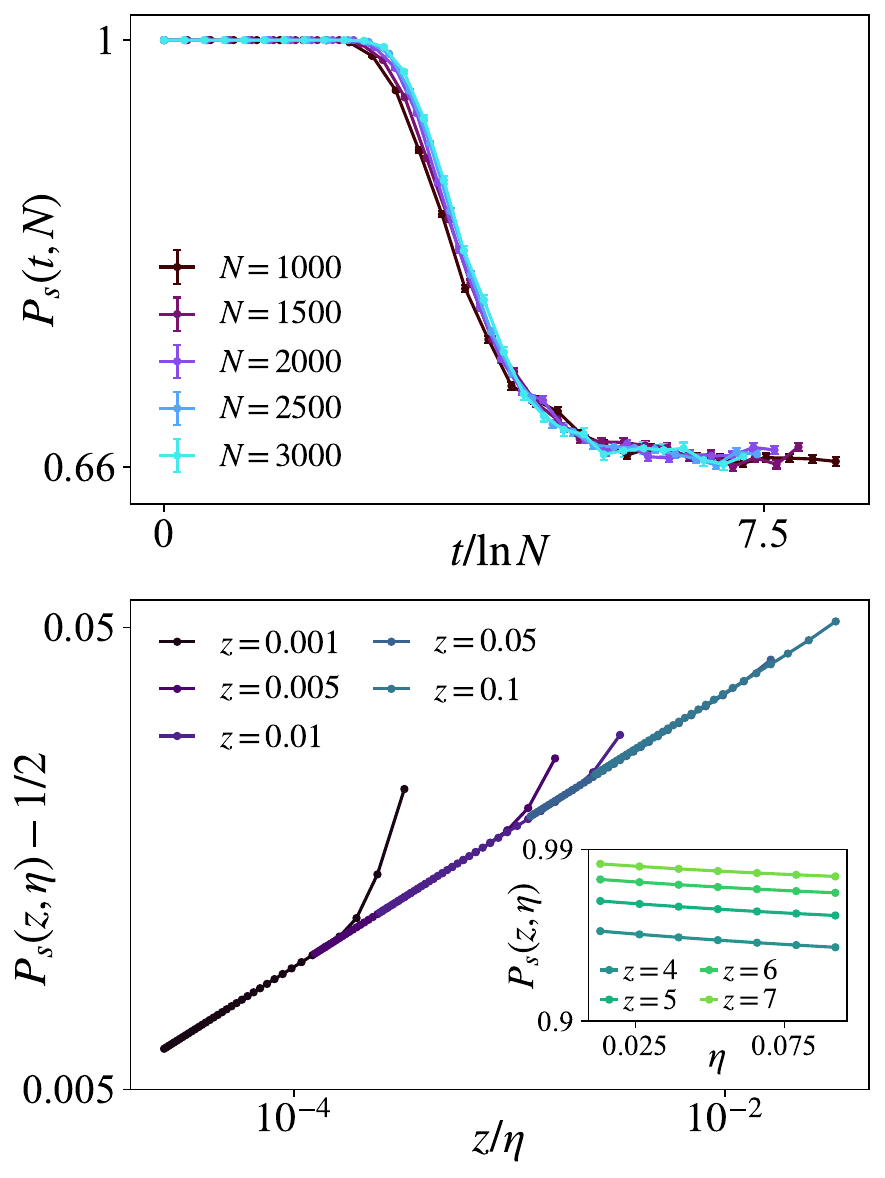}
    \caption{ Top: onset of anti-concentration in finite-depth Clifford circuits.  The decoder success probability $P_s(t, N)$ decays from one to the expected value for random Clifford circuits $P_s(t \rightarrow \infty, N) =2/3$ as the depth $t$ of the circuit is increased.  The curves for different system sizes collapse neatly when rescaling by $t/\ln(N)$, indicating that anti-concentration sets in on logarithmic timescales.  
    Bottom: The success probability of an RMPO displays $ z/\eta$ scaling with $\eta$ the rescaled noise strength in the small-$z$ (deep) limit, and is independent of $\eta$ in the large-$z$ (shallow) limit (inset).  The scaling in these two regimes is consistent with behavior seen in the linear XEB.}
    \label{fig:anti-concentration}
\end{figure}

We now consider candidate states prepared from finite-depth random circuits from initial computational basis states $\ket{\psi_\alpha(0)} = \ket{\vec{s}_\alpha}$ where $\vec s_0 \neq \vec s_1$.  In this setup, the two candidate states are perfectly distinguishable with $P_s=1$ at time $t=0$.  As successive gate layers are applied, the Born probability distribution is deformed from $P(w) = \frac{1}{D}\delta(w-1) + (1-\frac{1}{D})\delta(w)$ to the Porter-Thomas distribution $P(w) = De^{-Dw}$ in the deep-circuit limit, where a decoder should succeed with $P_s=3/4$ for a Haar-random circuit.  During this process, the distribution of measurement outcomes spreads out over the full computational basis-- this physical property is often referred to as \textit{anti-concentration}.  Given its relevance to computational hardness arguments \cite{aaronson2011computational,Bremner_2016,aaronson2016complexity,hangleiter2018}, detecting the onset of anti-concentration is a problem of practical significance.

The standard probe of anti-concentration is the linear cross-entropy benchmark (XEB).  For an initial pure state $\ket{\psi_\alpha(t=0)}$ that is then acted upon by a depth-$t$ circuit $U$ subject to some noise channel $\Ep$, where $\rho^\alpha(U) = U \ketbra{\psi_\alpha(0)}{\psi_\alpha(0)} U^\dag$ and $\rho_\Ep^\alpha(U)$ represents the noisily-evolved copy of the state, the linear XEB may be expressed as: 
\begin{equation}
    \begin{aligned}
        \mathrm{XEB} = D \sum_{\vec m } p\big(\vec m | \rho^\alpha(U) \big)  p\big( \vec m| ~\rho_\Ep^\alpha(U) \big) -1.
    \end{aligned}
\end{equation}
Heuristically, the linear XEB measures the correlation between the ideal distribution of Born probabilities of the prepared state $\rho^\alpha(U)$ and the distribution encoded in the state $\rho^\alpha_\Ep(U)$ produced by the noisy circuit.  Due to its relationship to the second moment $I_2$ of the Born probability distribution \cite{Bremner_2016, Hangleiter_2025}, the linear XEB is capable of probing anti-concentration properties \cite{Lami2025, Sauliere2025, sauliere2026}.  The XEB is also notably used in the context of RCS benchmarking and fidelity estimation \cite{boixo2018characterizing,arute2019quantum,ware2023sharp,morvan2024phase}; however, its use in benchmarking has come under scrutiny in recent years due to its vulnerability to classical spoofing attacks \cite{aaronson2020spoofing,barak2021spoofing,gao2024limitations}.

We propose that the decoder success probability may be a compelling alternative to the linear XEB for the purpose of probing anti-concentration properties.  The principal advantage of $P_s$ over the XEB is that it is dependent upon the full distribution of Born probabilities rather than only its second moment, which may make it a more robust probe of anti-concentration than the linear XEB --- a similar argument was recently presented in Ref. \cite{bentsen2026} in the context of robustness against spoofing.  Furthermore, the classifier confers an additional advantage in that the success probability is a more conceptually straightforward object from an information-theoretic perspective than the linear XEB.  

In order to demonstrate the viability of the decoder as an alternative to the XEB probe, we present numerical evidence below that confirms that $P_s$ exhibits several key scaling properties relevant to the detection of anti-concentration that are known to appear in the XEB.

\subsubsection{Anti-concentration onset at logarithmic circuit depths}

A key result is that even as deep circuits are required to prepare Haar-random states, random unitary circuits induce anti-concentration on timescales logarithmic in system size \cite{Dalzell_2022}.  To verify the presence of this scaling in the decoder success probability, we numerically study $P_s$ as a function of depth for states $\ket{\psi_\alpha}$ prepared from noiseless random Clifford circuits.  Here, at time $t=0$, $\ket{\psi_\alpha(t=0)}=\ket{\vec{s}_\alpha}$ for $\alpha=0,1$ where $\ket{\vec s_\alpha}$ is a randomly-selected computational basis state.  The condition $\vec s_0 \neq \vec s_1$ does not need to be imposed by hand in the large-$N$ limit at $t=0$: since there are $D=2^N$ states, $P[\vec{s}_0 = \vec{s}_1] = \frac{1}{D} \to 0$ and so $P_s(t=0)=1$.  We then apply a brickwork unitary circuit composed of $t$ layers of 2-qubit gates such that $U(t) = \prod_{\tau=1}^t U_\tau $, where, with periodic boundary conditions
$
U_\tau =
\bigotimes_{j=0}^{N/2-1}
V_{2j+r_\tau,\,(2j+1+r_\tau)\bmod N}(\tau)$ with $r_\tau=\tau\bmod2$,  and each two-qubit gate is chosen uniformly from the two-qubit Clifford group.

The numerical results are displayed in Fig.~\ref{fig:anti-concentration}.  The success probability drops from $P_s(t=0)=1$ in the shallow-circuit limit corresponding to distinguishing  product states, down to $P_s(t \gg 1) \rightarrow 2/3$ in the deep-circuit limit (Fig. \ref{fig:anti-concentration}).  The $2/3$ asymptotic value follows from a short calculation-- see Appendix \ref{app:clifford} for further information.  Furthermore, the curves display $t \sim \ln(N)$ scaling, consistent with analytical predictions \cite{Dalzell_2022}.

\subsubsection{Noisy finite-depth scaling} \label{subsec:noisy-finite}

We would now like to extend our analysis to include different scaling regimes in noisy finite-depth circuits. It is well-established that the linear XEB undergoes a phase transition from a low-noise regime in which it closely approximates the state fidelity to a high-noise regime in which this correspondence breaks down \cite{ware2023sharp, morvan2024phase}.  Here we numerically demonstrate that the classifier exhibits the same scaling as the linear XEB in both of these regimes, which therefore indicates that it is sensitive to the same transition. 

Following the approach of \cite{Lami2025,Sauliere2025,sauliere2026}, we represent a noisy circuit by using a random matrix product operator (RMPO) of $N$ qubits of bond dimension $\chi^2 \sim e^{2t}$ as a proxy for states prepared with noisy random circuits of depth $t$.  Circuit depth may be tuned relative to system size by adjusting the parameter $z = N(d-1)/(d \chi)$, where here $d=2$ is the local Hilbert space dimension; the limit $z \gg \eta$ probes the shallow-circuit regime while $z \ll \eta$ probes the deep-circuit regime, where $\eta = qN$ is a rescaled error rate and $q$ is the noise strength of the channel.  While an exact expression for the distribution of Born probabilities is not analytically tractable, the forms of its moments $I_k =\sum_{\vec{m}} p(\vec{m})^k$ are known exactly \cite{sauliere2026}.  Again as in Ref. \cite{sauliere2026}, the full distribution of Born probabilities $P(u = D p(\vec{m}); z, \eta)$ may be approximated from its moments using an Edgeworth expansion, which allows for the numerical evaluation of $P_s(z, \eta)$ across a range of both parameters (Fig. \ref{fig:anti-concentration}).  

In the deep-circuit regime ($z \ll \eta$), we find that the decoder success probability approaches $P_s=1/2$ as $\eta$ increases at fixed $z$, with the curves collapsing as a function of $z/\eta$. To leading order in the shallow-circuit regime ($z \gg \eta$), the success probability is approximately independent of $\eta$ (Fig.~\ref{fig:anti-concentration}). The scaling of the success probability in these different regimes is consistent with the results reported for the linear XEB in Ref. \cite{sauliere2026} and the transition reported in Ref. \cite{ware2023sharp}.

\section{Discussion}

Despite their local indistinguishability, global snapshots of random quantum states contain a surprising amount of information, provided that they are paired with a classical model of all possible candidate states that can be efficiently queried.  With access to just a single bitstring and employing an optimal Bayesian classifier as a decoder, two random candidate states may be distinguished from each other with probability $P_s=3/4$; the decoder success probability may be raised or lowered by increasing the number of snapshots $m$ or number of candidate states $k$, respectively.  We present a theory of a phase transition in decodability as a function of the variable $x=m/\ln k$ and derive the threshold value $x_c=1/(1-\gamma)$, where $\gamma$ is Euler's constant, in the limit of large $k$ and $m$.  While the decodability transition follows naturally from standard classical channel coding arguments, in the state distinguishability context our result strikingly indicates the number of random states that may be distinguished is exponential in shot number.  We extend our analysis to variants of the $k=2$, $m=1$ distinguishability game and present numerical results for the behavior of the classifier when the candidate states are phase-random, noisy, or prepared with finite-depth circuits. We also generalize the threshold analysis to noisy observations decoded using ideal candidate probabilities.

Given the surprising amount of information contained in global snapshots, optimal Bayesian classifiers likely have many applications across quantum sampling experiments.  One potential application of the distinguishability game briefly explored here is as an alternative to the linear cross-entropy benchmark as a probe of anti-concentration -- see Ref.~\cite{bentsen2026} for a similar discussion from the spoofing perspective -- but other potential applications should be explored in future work. 

\textbf{{\textit{Acknowledgments---}}}  We thank Jacopo De Nardis, Tibor Rakovszky and Curt von Keyserlingk for insightful discussions and useful suggestions. We acknowledge support from the Swiss National Science Foundation (grant 10008234, CM and RV), the Foundation for the University of Geneva (RV), and  through the Co-design Center for Quantum Advantage (C2QA) under contract number DE-SC0012704 (SG). We acknowledge the use of generative AI tools for assistance with calculations and preparation of numerical scripts, along with proofreading and polishing the manuscript.

\appendix

\section{Phase-random states} \label{app:phase-random}

\begin{figure}
    \centering
    \includegraphics[width=0.85\linewidth]{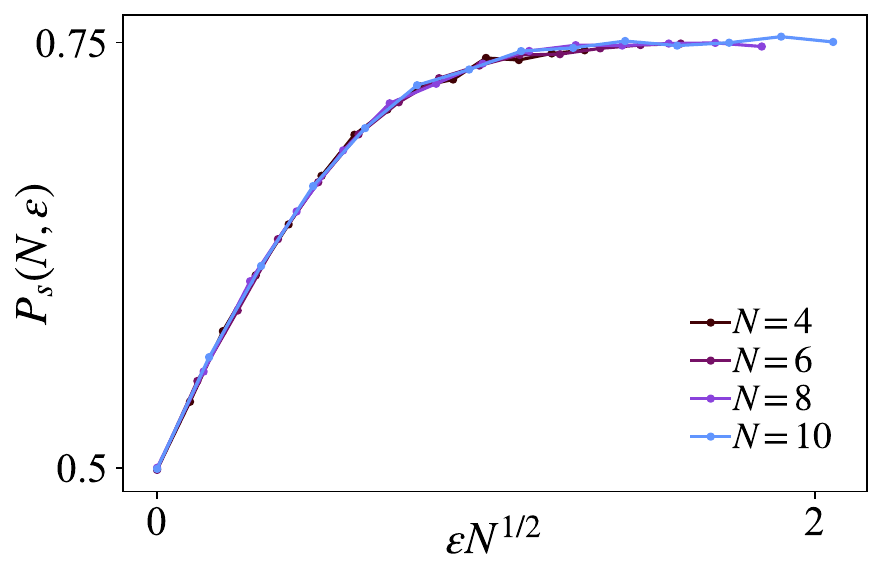}
    \caption{Success probability of phase-random states measured in a basis defined by the rotation $U = \bigotimes_j e^{-i \epsilon X_j}$, where the constant $\epsilon$ determines the size of the rotation.}
    \label{fig:phase-random}
\end{figure}

The $\epsilon N^{1/2}$ scaling arises from properties of the Born probability distribution in the new basis.  Without access to the full form of this distribution, we study the analytical form of the second moment of the distribution of Born probabilities in the new basis in order to understand the scaling variable.  For $\ket{\psi_\alpha} = \frac{1}{\sqrt D}\sum_{\vec s} e^{i \theta_{\vec s,\alpha}}\ket{\vec{s}}$ with $\alpha=0,1$ and random $\theta_{\vec s,\alpha} \in [0,2\pi )$, the Born probabilities in the new basis are given by $p(\vec m | \psi_\alpha, \epsilon) = \frac{1}{D}|\sum_{\vec s} c_s e^{i \theta_{\vec m \oplus \vec s}}|^2$, with $c_s = (\cos \epsilon)^{N-|\vec s|} (i \sin \epsilon)^{|\vec s|}$.  Rearranging terms gives the following expression for the second moment:
\begin{equation}
    \mathbb E [p(\vec m | \psi_\alpha, \epsilon)^2]= \frac{1}{D^2} \Big[2 \Big(\sum_{\vec s} |c_{s}|^2 \Big)^2 -\sum_{\vec s} |c_s|^4\Big].
\end{equation}
The first sum in the above expression evaluates to $\sum_{\vec{s}} |c_s|^2=1$, while the second sum simplifies to $\sum_{\vec s} |c_s|^4 = (1 - \frac{1}{2} \sin^2(2 \epsilon))^N$.  After taking the small-$\epsilon$ limit $N \epsilon^2 \ll 1$, we obtain:
\begin{equation}
    \begin{aligned}
        \mathbb{E}[p(\vec m | \psi_\alpha, \epsilon)^2] \approx \frac{1}{D^2}(1+2N\epsilon^2),
    \end{aligned}
\end{equation}
which features the expected $\epsilon \sim N^{-1/2}$ scaling.  Interestingly, although both moments scale as $D^{-2}$, this second moment does not generally coincide with the Haar value
$\mathbb{E}[p(\vec m|\psi_\alpha)^2]_{\mathrm{Haar}}
=\frac{2}{D(D+1)}$.  The fact that the ensemble does not form an exact state 2-design is consistent with the intuition that the mechanism underpinning the high success probability has more to do with heavy-output generation rather than the specifics of the Porter-Thomas distribution.

\section{Decoder scaling with random coherent errors} \label{app:coherent}

Here we derive the scaling of the decoder success probability in the presence of random coherent errors (Sec. \ref{subsec:coh}).  The success probability can be derived by studying the Born probabilities averaged over $\epsilon_j$ drawn iid from a Gaussian distribution with variance $\sigma^2$.  In the computational basis, the probabilities are given by:
\begin{equation}
    \begin{aligned}
        p'(&\vec m| \psi_\alpha) = | \bra{\vec m} \bigotimes_j e^{-i \epsilon_j X_j} \ket{\psi_\alpha} |^2 .
    \end{aligned}
    \label{eq:noisyprobs}
\end{equation}
Expanding $e^{-i\epsilon_jX_j}
=\cos\epsilon_j\,\mathbf{1}_j-i\sin\epsilon_j\,X_j$,
cross terms between distinct flip patterns vanish upon averaging,
because they contain a factor
$\mathbb{E}_{\epsilon_j}[\sin\epsilon_j\cos\epsilon_j]=0$
at a site where the patterns differ.
The remaining diagonal terms are evaluated using
$\mathbb{E}_{\epsilon_j}[\cos^2\epsilon_j]
=(1+e^{-2\sigma^2})/2$
and
$\mathbb{E}_{\epsilon_j}[\sin^2\epsilon_j]
=(1-e^{-2\sigma^2})/2$. We have:
\begin{equation}
    \begin{aligned}
        p'(\vec m | \psi_\alpha) = \sum_{\vec s } c_s ~| \braket{\vec m \oplus \vec{s}|\psi_\alpha}|^2,
    \end{aligned}
    \label{eq:noisyaveraged}
\end{equation}
where the prefactors are defined as $c_s = ( \frac{1 + e^{-2 \sigma^2}}{2} )^{N-|\vec s|}( \frac{1 - e^{-2 \sigma^2}}{2} )^{|\vec s|}$.  The success probability of a decoder that uses the original Born probabilities as inputs but measures the state after the application of coherent errors is given by:
\begin{equation}
    \begin{aligned}
        &P_s(\sigma) = \frac{1}{2}\sum_{\vec m,\alpha=0,1} p'(\vec m| \psi_\alpha) \Theta\big( p(\vec m| \psi_\alpha) - p(\vec m| \psi_{\alpha \oplus 1})\big) 
    \end{aligned}
\end{equation}
which evaluates to $\frac{3 c_0}{4} + \sum_{\vec s \neq \vec 0} \frac{c_s}{2} \big[\sum_{\vec m,\alpha=0,1} p(\vec m \oplus \vec s| \psi_\alpha) \Theta( p(\vec m| \psi_\alpha) - p(\vec m| \psi_{\alpha \oplus 1}))\big]$.  Since the contributions of both $\alpha=0,1$ in the bracketed term are symmetric, the sum over both $\alpha$ cancels the leading factor of $1/2$.  The remaining expression is then in terms of three probabilities drawn independently from the Porter-Thomas distribution, and so $\sum_{\vec m} p(\vec m \oplus \vec s| \psi_\alpha) \Theta( p(\vec m| \psi_\alpha) - p(\vec m| \psi_{\alpha \oplus 1})) = \big( \int_0^\infty du \,e^{-u} \int_{0}^u du' \,e^{-u'} \big) \big(\int_0^\infty du \,ue^{-u} \big)$.  Since this expression approaches the value of $1/2$ in the $D \rightarrow \infty $ limit, the success probability of the noisy decoder then is given by the following:
\begin{equation}
    P_s(\sigma)  = \frac{1}{2} + \frac{1}{4}\Big( \frac{1 + e^{-2 \sigma^2}}{2} \Big)^N.
\end{equation}
The scaling behavior as $P_s(\sigma)$ approaches 1/2 is then $ P_s(\sigma) - \frac{1}{2} \sim e^{-\sigma^2 N}$ for $\sigma \ll 1$, consistent with the numerically-obtained scaling collapse shown in Fig. \ref{fig:noise} in the main text.

\section{Decoder success probability in deep Clifford circuits} \label{app:clifford}

Here we derive the deep-circuit Clifford limiting value $P_s(t \rightarrow \infty)=2/3$.  This value follows straightforwardly from the observation that any two states $\ket{\psi_0}, \ket{\psi_1}$ prepared with the same Clifford circuit will have support over the same $2^g$ number of basis states at a specific circuit depth $t$, where $g$ is known as the participation entropy.  For each fixed value of $g$, the probabilities $p(\vec m|\psi_{0,1})$ may only take the values $0$ or $1/2^g$ and are distributed according to $P_g(w) = (1-2^g/D) \delta(w) + (2^g/D) \delta(w-1/2^g)$, where $D=2^N$ is the Hilbert space dimension and $w=p(\vec m| \psi_\alpha)$.  Using $P_g(w)$ as the appropriate Born probability distribution for two states with a fixed value of $g$ yields
\begin{equation}
    \begin{aligned}
        P_s&(g) = 1-2^{g-1}/D.
    \end{aligned}
    \label{eq:shallow}
\end{equation}

The total success probability is given by $P_s = \sum_gP_s(g)P(N,g)$, where $P(N,g) = \binom{N}{g}_2 \frac{2^{g(g+1)/2}}{(-2;2)_N}$ \cite{magni2025}, which approaches $P_s=2/3$ in the large-$D$ limit.

Note that the $P_s(t \gg1 ) = 2/3$ result only holds for the case when the two states are prepared from different computational basis states with the same circuit; the reason for this is that states prepared from different computational-basis states with the same Clifford circuit have the same participation entropy whereas two entirely random Clifford states will generically have different participation entropies in the large-$D$ limit.  Repeating the above calculation with different participation entropies $g_0, g_1$ gives $P_s(g_0,g_1) = 1-2^{\min(g_0,g_1)-1}/D$; performing the sum $P_s = \sum_{g_0,g_1} P(N,g_0) P(N,g_1) P_s(g_0,g_1)$ yields a value $\approx 0.7458$, slightly below the $3/4$ asymptotic value obtained for the Haar case.

\bibliography{references}
\bibliographystyle{unsrt}

\end{document}